# Investigations of thermometric characteristics of $p^+$-$n$ - type GaP diodes

**V.A. Krasnov, Yu.M. Shwarts, M.M. Shwarts, D.P. Kopko,**

**A.M. Fonkich, S.Yu. Yerochin, S.V. Shutov, N.I. Sypko**

V. Lashkaryov Institute of Semiconductor Physics NAS Ukraine

41 prospekt Nauky, Kyiv, Ukraine, 03028
Tel/fax: +38 (044) 525-74-63,
e-mail: shwarts@isp.kiev.ua, shutov_sv@mail.ru

**Abstract**

*The technique of obtaining of $p^+$-n-type gallium phosphide diode epitaxial structures from liquid phase was developed as well as pilot samples of diode temperature sensors were fabricated based on them. Thermometric and current-voltage characteristics of the test diodes were measured in the temperature range of 80÷520K and their basic technical characteristics were determined. An availability of application of the structures developed as sensing elements of high-temperature heat sensors was shown.*

## Introduction

The need of creation of a new generation of precise and replaceable temperature sensors based on wide band-gap diodes is caused by requirements of extreme-temperature electronics, so the investigations in this area are in progress now [1]. Gallium phosphide is known as a semiconductor material for fabrication of light-emitting diodes for visible spectrum range, particularly for red and green glow. At the same time wide band-gap GaP diode structures ($E_g \approx$ 2.25 eV at T=300K) are promising for building devices of high-temperature electronics including high-temperature heat sensors [2-4]. Therefore development of methods of obtaining of GaP-based epitaxial structures for creation of sensing elements of high-temperature sensors is an actual problem for sensor electronics.

The purpose of the present work is to work out and to test the technology of fabricating of GaP-based diodes of $p^+$-$n$-type with design and technological parameters that ensure their application in high-temperature thermometry.

## Methodology and experiment

The process of liquid phase epitaxial growth of n-$p^+$-GaP structures is described in [5]. Contact layers of Ni alloy were deposited by the methods of thermal and ion sputtering with consequent deposition of Au or Ag. Planar operations were completed by forming of mesa structure using a masking technique as well as photolithography and selective etching. After cutting the wafers into chips using standard technology of semiconductor device manufacturing, operations of chip thermal compression over a kovar holder, chip terminals welding as well as encapsulation into a case of thermosetting plastic were performed.

The thickness of epitaxial layers was controlled by making a cleavage or a spherical slice. The accuracy of the film thickness measurements made ~ 0.3 μm. Determination of charge carrier concentration and the character of its thickness distribution were conducted by capacity-voltage characterization of the structures. Layer-by-layer etching was used to characterize a thickness distribution of the concentration. Area homogeneity of doping of GaP epitaxial structures was controlled by breakdown voltage $U_b$ of test diodes with Schottky barrier. Dispersion of the $U_b$ magnitude showed non-homogeneity of the doping. For studying of thin-film $p^+$-$n$-structures of GaP the structures with doping non-homogeneity below 5% were selected. To evaluate a dislocation density $N_D$ in $n$- and $p^+$- layers of GaP a metallographic method was applied. Diameter and thickness distribution of $N_D$ in epitaxial film was determined during layer-by-layer etching of GaP layers in polishing etch with subsequent selective etching of the layer and dislocation exposure by decoration technique. For the films investigated $N_D \approx (1 \div 5)\cdot 10^4$ cm$^{-2}$.

Charge carrier concentration and its mobility in $n$- and $p^+$- layers of GaP epitaxial structure was defined by Hall effect measurements at two temperatures (77.4K and 300K) using technological satellite substrates.

Thermometric characteristics (TMC) of the diodes for three operation currents 1, 10 and 100 μA were measured in the range of 78-520K. The accuracy of the operation current maintenance was not worse than ±0.1%. General absolute error of the temperature measurements did not exceed ±0.03K.

Forward current-voltage characteristics (CVC) of the diodes were measured using an automatic apparatus in the interval of currents $10^{-11}$ - $10^{-2}$A and temperature range of 77.3÷473K. Temperature maintenance error did not exceed ±0.1K.

## Results and discussion

Results of measurements of thickness, electrophysical and structural characteristics of epitaxial $p^+$- $n$- structures of GaP are presented in Table 1.

Table 1.

Parameters of $p^+$- $n$ structures of GaP investigated obtained by liquid phase epitaxy

| Layer | Layer thickness, μm | Charge carrier concentration, см$^{-3}$ | Hall charge carrier mobility μ, cm$^2$·V$^{-1}$s$^{-1}$ | | Dislocation density, cm$^{-2}$ |
|---|---|---|---|---|---|
| | | | 77.4 K | 300 K | |
| $n$-GaP | $d_n \approx (3\div7)$ | $n \approx (5\cdot10^{14}\div2\cdot10^{15})$ | 1100÷1220 | 230÷270 | $N_{dn} \leq 3\cdot10^4$ |
| $p^+$-GaP | $d_{p+} \approx (10\div15)$ | $p^+ \approx (2\div3)\cdot10^{18}$ | - | - | $N_{dp} \leq 3\cdot10^4$ |

To carry out metrological studies a set of ten gallium phosphide diodes with the following parameters was formed: $d_n \approx (5\pm1)$μm, $d_{p+} \approx (12\pm2)$μm, $n \approx (5\div8)\cdot10^{14}$ cm$^{-3}$, $p^+ \approx (2\div3)\cdot10^{18}$ cm$^{-3}$, μ (300 K) ≈ 250 cm$^2$·V$^{-1}$s$^{-1}$, $N_{dn} \approx 8\cdot10^3$ cm$^{-2}$, $N_{dp} \approx 1.5\cdot10^4$ cm$^{-2}$.

Metrological studies showed that at room temperature the forward voltage magnitude dispersion from the rated value in the set of ten test thermodiodes lied in the range of ± 4 mV.

Typical TMC and temperature dependency of the test diodes sensitivity at three values of operation current are presented in fig.1 and 2 correspondingly. It should be noticed that at the temperature of 500K voltage drop across the structure makes 1.05÷1.3V depending on the current magnitude. The studies showed that reproducibility of temperature properties of the diodes during 5 months made ~± 2% .

As it could be seen in fig.2, for the current of 1 μA TMC in the temperature range of 78-520K is distinguished by its high linearity with sensitivity ~ -2.5 mV/K. For the current of 10 μA a deviation from quasilinearity of TMC and subsequent sensitivity increase are observed at the temperatures below ~170K, and for 100 μA – below ~250K.

Fig.3 shows CVC of the diodes measured in the range of 77÷463K. Ideality factor vs. current dependencies were determined from the analysis of CVC (fig. 4). It can be seen that for the current of 1μA in the whole temperature range investigated a recombination mechanism of charge carrier transport with ideality factor close to 2 is dominant. For the currents of 10 and 100 μA domination of a recombination mechanism remains only at the temperatures above 160K and 250K respectively. The value of activation energy found from dependencies *I(T)* at the constant diode voltage (fig.5) equals 1.2eV, which corresponds to the half band-gap of GaP.

## Conclusion

The technique of obtaining of epitaxial $p^+$-$n$ -structures based on GaP having an improved reliability and minimal dispersion of the forward voltage drop was worked out. Thermometric characteristics of the diodes developed were determined and availability of their application as sensitive elements of high-temperature thermal sensors was shown.

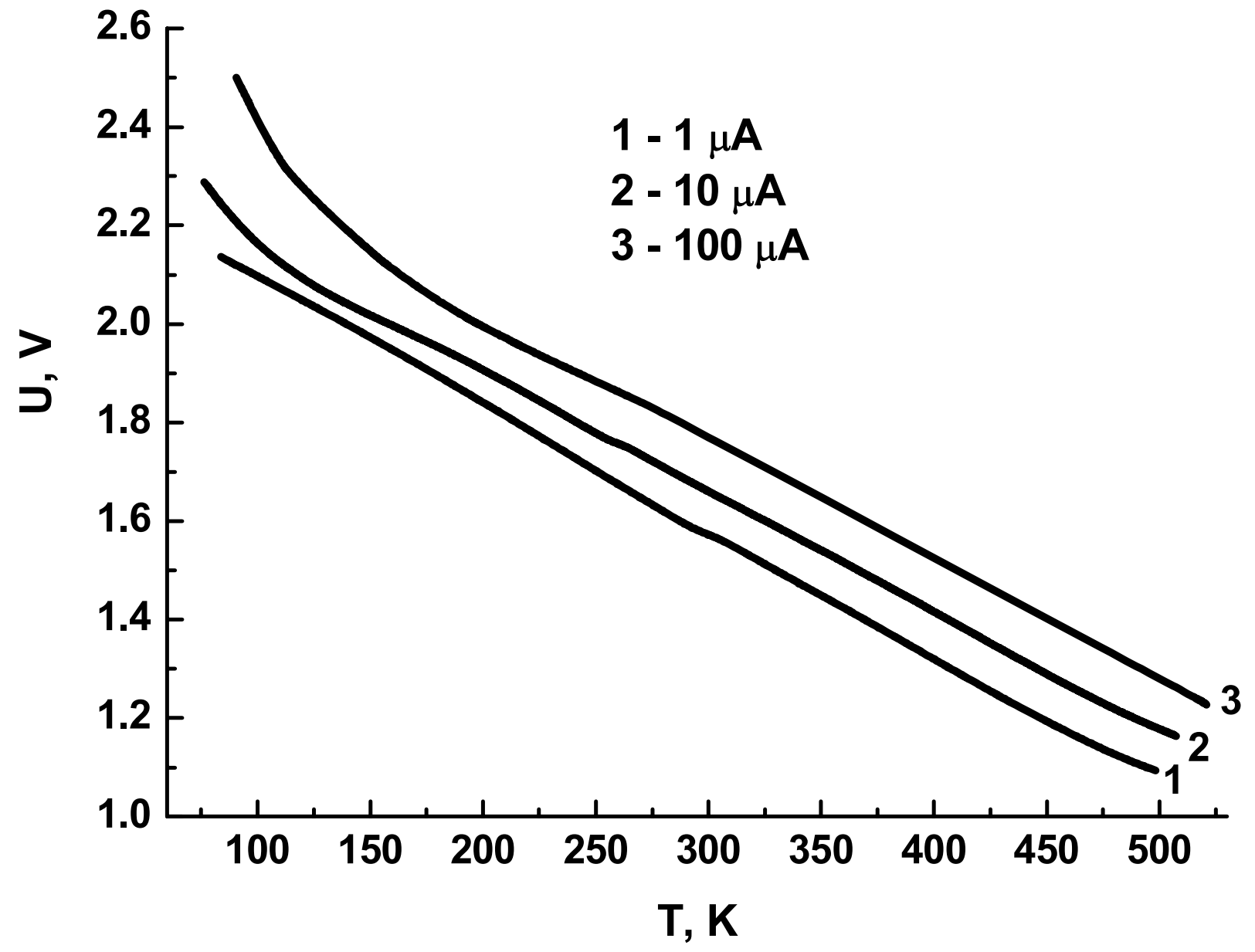

Fig.1. Thermometric characteristics of *n-p*$^+$-GaP diodes for three values of the operation current: 1, 10 and 100 μA.

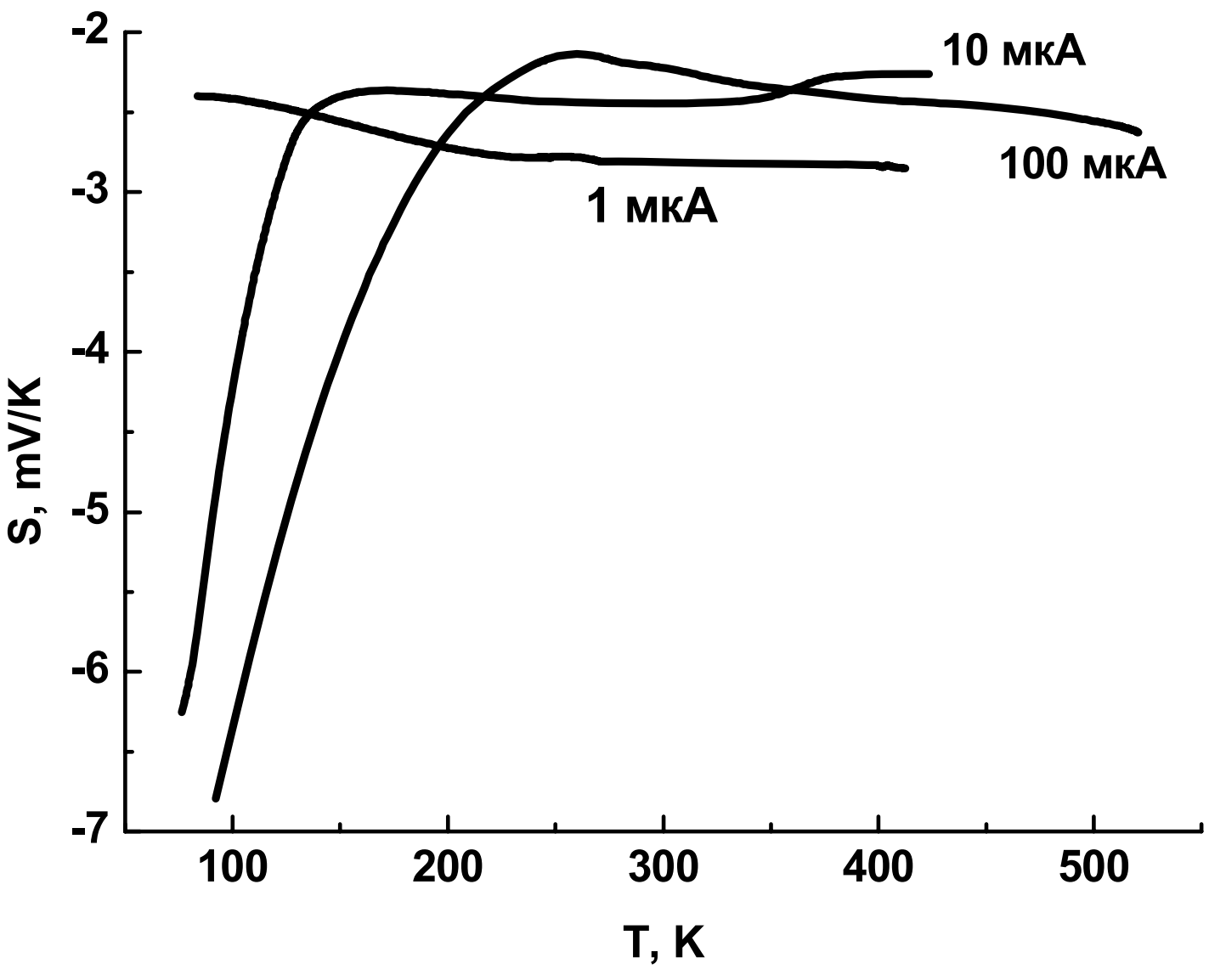

Fig.2. Sensitivity of $n$-$p^+$-GaP diodes for the operation current: 1, 10 and 100 μA.

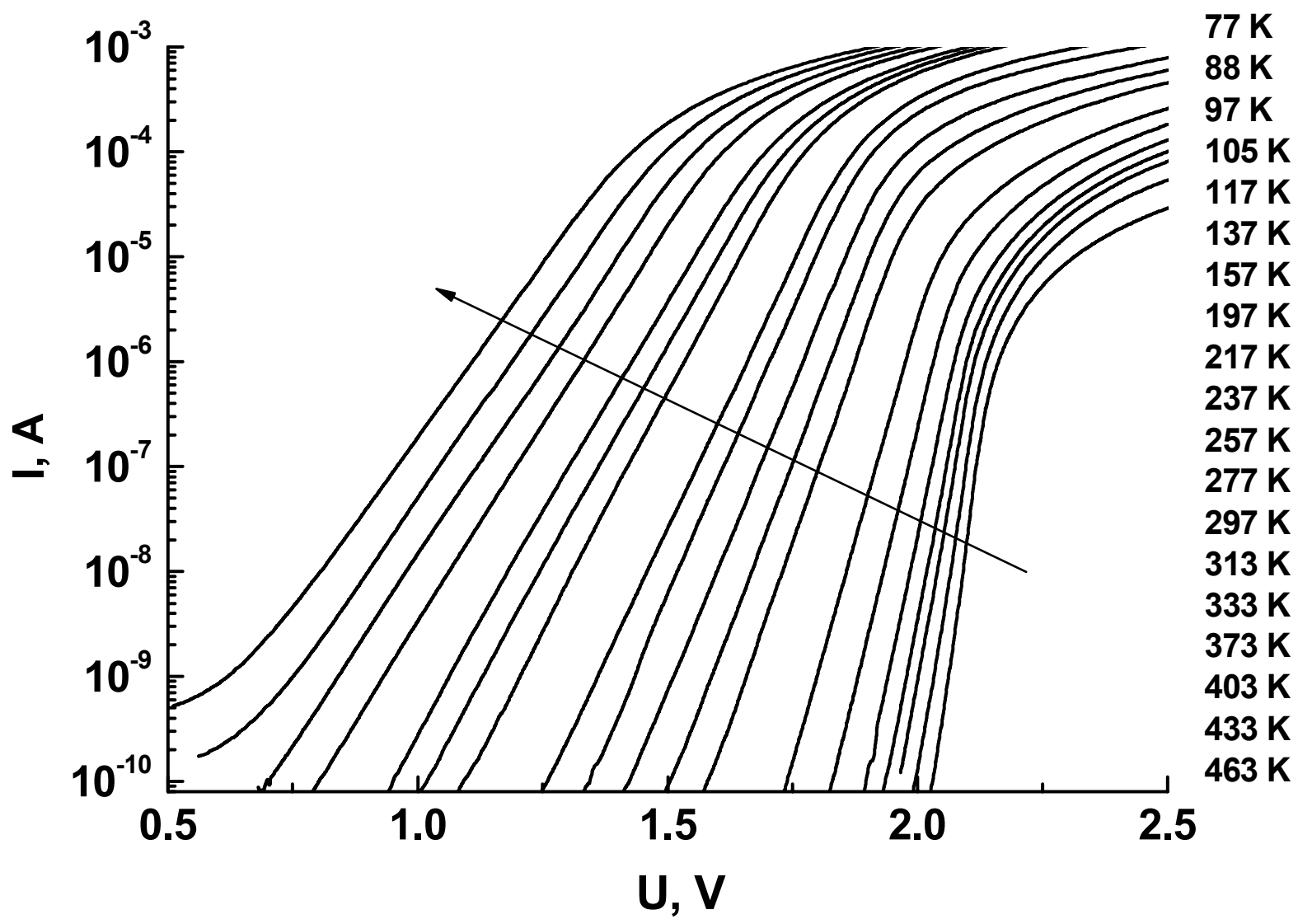

Fig.3. Current-voltage characteristics of GaP diodes measured in the temperature range of 77-463 K. The arrow shows a direction of the temperature increase.

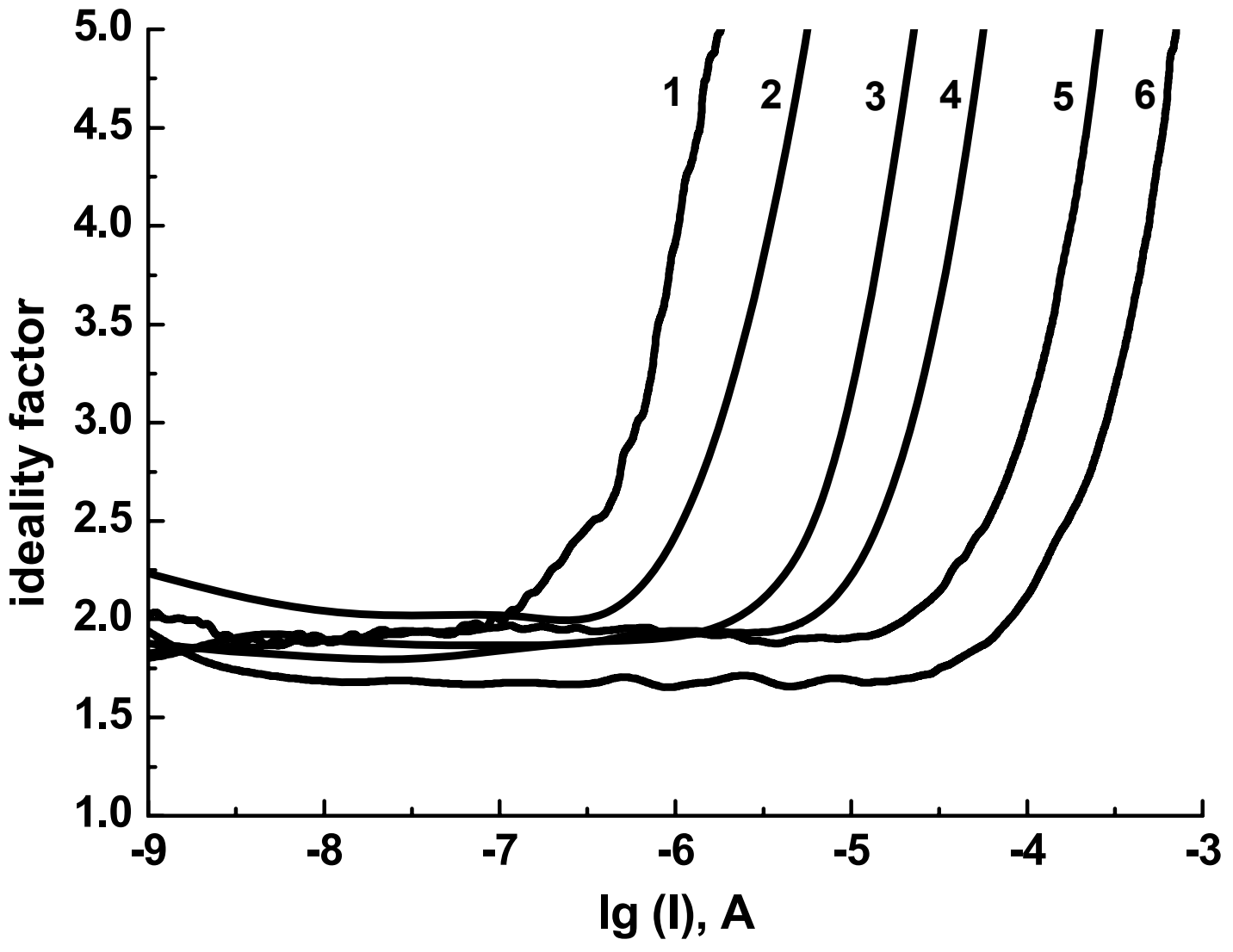

Fig.4. Ideality factor vs. current dependencies for GaP diodes investigated measured at different temperatures, K: 1 - 77, 2 - 129, 3 - 160, 4 - 208, 5 - 250, 6 – 471.

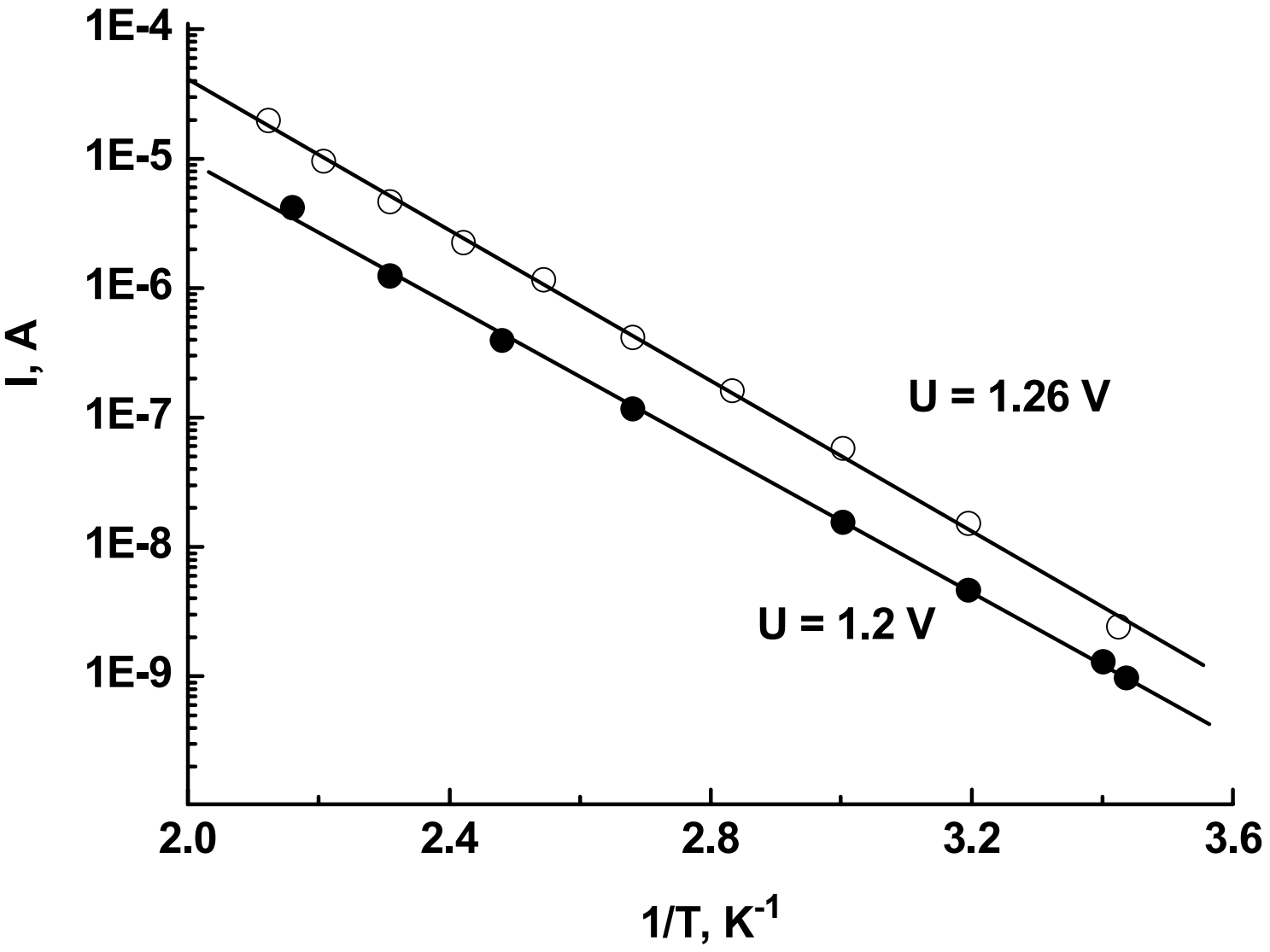

Fig.5. Current vs. $1/T$ dependencies at U=1.2V and U=1.26V for GaP diodes investigated.